\documentclass[a4paper,11pt]{article}

\usepackage{jheppub} 
\usepackage[T1]{fontenc}
\usepackage{graphicx}

\newcommand{\bg}{\mbox{\boldmath $\gamma$}}
\newcommand{\bt}{\mbox{\boldmath $\theta$}}

\newcommand{\bx}{\mbox{\boldmath $x$}}

\title{\boldmath Correspondence between causality in flat Minkowski spacetime and entanglement in thermofield-double state: Hessian-geometrical study}

\author{Hiroaki Matsueda}
 
\affiliation{Sendai National College of Technology, Sendai 989-3128, Japan}

\emailAdd{matsueda@sendai-nct.ac.jp}

\abstract{
We examine the Hessian potential that derives the flat Minkowski spacetime in $(1+1)$-dimension. The entanglement thermodynamics by the Hessian geometry enables us to obtain the entanglement entropy of a corresponding quantum state by means of holography. We find that the positivity of the entropy leads to the presence of past and future causal cones in the Minkowski spacetime. We also find that the quantum state is equivalent to the thermofield-double state, and then the entropy is proportional to the temperature. The proportionality is consistent with previous holographic works. The present Hessian geometrical approach captures that the causality in the classical side is converted into quantum entanglement inherent in the thermofield dynamics.
}


\begin{document}

\maketitle

\flushbottom

\section{Introduction}

Similarity of black hole physics with quantum entanglement is a common keyword in interdesciplinary physics fields. In this short note, we are going to show an explicit example of this similarity by means of holographic transformation between them like the AdS/CFT correspondence~\cite{Maldacena1,Maldacena2}. Namely, we start with a causal surface on the flat Minkowki metric, and we construct a dual quantum field theory with the help of the Fisher information metric. Although the example might be too simple, the correspondence enables us to facilitate information-geometrical understanding of such transformation. We will apply the Hessian geometry to this problem~\cite{Shima}, since we have recently been developing this approach and have realized that this is quite powerful to understand fundamental notions of the AdS/CFT correspondence~\cite{Matsueda1,Matsueda2}. This work will make this situation much better, and I am going to provide readers with some positive feeling for applying the Hessian geometry to several problems related to the AdS/CFT correspondence. Up to now, I have been focusing on the hyperbolic space, and thus we slightly shift our mind to examine the flat space holography. The holographic entropy calculation in the flat spacetime has been performed in Ref.~\cite{Li}, where it is argued that the reduced density matrix gives the maximal entropy and the correlation functions become trivial. We will argure that the present result also supports this feature.

Motivated by the above current status, we examine the Hessian potential that is converted to the flat Minkowski spacetime with the help of the Fisher information metric. This Hessian potential can be also transformed into the entanglement entropy in the quantum side by using the entanglement thermodynamics. We will find that the entropy is maximal and obeys the volume law scaling. We will also find that the dual quantum field theory is described by the thermofield dynamics (TFD)~\cite{TFD}. The result suggests that the entanglement nature between dual Hilbert spaces in TFD is basically equal to the causality in the classical side.

The most important thing would be about the coordinate transformation as we have already mentioned in my previous papers~\cite{Matsueda1,Matsueda2}. When we consider some holographic transformation, the original model parameters in the quantum side are converted to the spacetime coordinates in the classical side. However, this is quite nontrivial transformation, and actually in the present case time $t$ and inverse temperature $\beta$ in TFD are converted to the spacetime coordinates in the classical side. This concept is not still fully recognized in usual holographic theories. We emphasize the importance of this nontriviality in this paper again.

The organization of this paper is as follows. In Sec.~II, we give a method of constructing the flat spacetime by the Hessian geometry, and discuss about the volume law for the entropy. In Sec.~III, we present how to transform the Minkowski metric to TFD and examine the nature of the thermal state emerging from the classical data. In Sec.~IV, we introduce the quantized Fisher metric to precisely describe how the time evolution of the quantum state is related to the Lorentzian signature in the classical side. The final section is devoted to the summary part.

\section{Causal cones in the classical side}

\subsection{Construction of Minkowski spacetime from Hessian potential}

In this paper, we rely on the basic notations in Refs.~\cite{Matsueda1,Matsueda2} except for replacement $\bt\rightarrow\bx$. The key quantity is the Hessian potential that derives both of the Fisher metric and the entanglement entropy. Let us start with the following bilinear form of the Hessian potential,
\begin{eqnarray}
\psi(\bx) = \frac{1}{2}\left(-t^{2}+x^{2}\right), \label{2d}
\end{eqnarray}
where we take $\bx=(x^{0},x^{1})=(t,x)$. The Fisher metric is defined by
\begin{eqnarray}
g_{ij}(\bx)=\partial_{i}\partial_{j}\psi(\bx),
\end{eqnarray}
for $i,j=0,1$. It is straightforward to derive the metric on the Minkowski spacetime $M^{1,1}$ from the above potential,
\begin{eqnarray}
ds^{2}=g_{ij}dx^{i}dx^{j}=\eta_{ij}dx^{i}dx^{j}=-dt^{2}+dx^{2}.
\end{eqnarray}
In general $D$-dimensional cases, we can extend Eq.~(\ref{2d}) to
\begin{eqnarray}
\psi(\bx)=\frac{1}{2}\eta_{ij}x^{i}x^{j}, \label{general}
\end{eqnarray}
and obtain the Minkowski spacetime $M^{1,D-1}$.

It is well-known in information geometry that the introduction of the Lorentzian signature induces complex probability distributions, and usually it brings some trouble in the theory. However, we will consider a situation that the distributions originate in coefficients in a wave function, and thus this severe condition may be relaxed. Later, we will discuss these points in more detail, and will confirm that the appearance of the complex variables does not bring any trouble.

\subsection{Volume law for the entanglement entropy and constraint for the Minkowski spacetime}

According to the entanglement thermodynamics based on the Hessian geometry~\cite{Matsueda1,Matsueda2,Casini,Blanco,Wong,Takayanagi2,Takayanagi3,Faulkner,Nima,Banerjee}, the entanglement entropy is related to the Hessian potential by the following equality,
\begin{eqnarray}
S(\bx)=\psi(\bx)-x^{i}\partial_{i}\psi(\bx),
\end{eqnarray}
and in the case of Eq.~(\ref{general}) we find
\begin{eqnarray}
S(\bx)=-\psi(\bx). \label{minus}
\end{eqnarray}
This condition is exactly satisfied for arbitrary dimension $D$. In the case of AdS/CFT previously examined~\cite{Matsueda1,Matsueda2}, $S$ and $\psi$ were identified with each other after the second derivative by the canonical parameters $\bx$. Thus, it is interesting that Eq.~(\ref{minus}) is rather different from the previous situation. We ensure the positivity of the entanglement entropy,
\begin{eqnarray}
S=\frac{1}{2}(t^{2}-x^{2})\ge 0. \label{wedge}
\end{eqnarray}
This constraint gurantees that we should only consider the inside of the past and future causal cones. According to previous works on similarity between the black hole physics (presence of a causal surface) and thermofield double~\cite{Matsueda3}, we easily imagine that the corresponding quantum state would be represented by the thermofield dynamics. In Eq.~(\ref{wedge}), the entropy value vanishes at the boundaries of the causal cones. Later we need to discuss how this feature is converted into the quantum side. If we start with the Hessian potential $\psi(\bx)=(1/2)\sum_{i}(x^{i})^{2}$ with use of real variables of canonical parameters $\bx$, we then obtain the Euclidean metric $g_{ij}(\bx)=\delta_{ij}$ but the entropy becomes negative: $S=-\psi=-(1/2)\sum_{i}(x^{i})^{2}\le 0$. To keep the positivity of the entanglement entropy, it is necessary to introduce the Lorentzian signature $\eta_{ij}$.

It is an interesting question whether the entanglement entropy obeys the volume law or not. To see this, it is helpful to evaluate the covariant derivative of the information flow $\nabla_{i}x^{i}$, since it's volume integral is equal to the entanglement entropy. The result for general $D$-dimensional case is as follows
\begin{eqnarray}
\nabla_{i}x^{i}=\partial_{i}x^{i}+\Gamma^{i}_{\;ij}\theta^{j}=-\frac{1}{2}g^{ij}\partial_{i}\partial_{j}\left(S-\psi\right)=D,
\end{eqnarray}
where the Christoffel symbol is described as
\begin{eqnarray}
\Gamma^{k}_{\;ij}=\frac{1}{2}g^{kl}\partial_{l}\partial_{i}\partial_{j}\psi(\bx).
\end{eqnarray}
The covariant derivative of $x^{i}$ takes a finite value, and thus the entanglement entropy obeys the volume law scaling in spite of the area law scaling in the case of the AdS/CFT correspondence. In particular, Eq.~(\ref{minus}) guarantees that the entropy tends to be maximally enhanced. Thus, if there is a quantum state dual to the causal cones, this quantum state is highly entangled one.

\section{Entanglement structure and wave function in the quantum side}

\subsection{Thermofield double state}

Originaly, the canonical parameters $\bx$ are defined from the eigenvalues of the reduced density matrix by the following representation
\begin{eqnarray}
\lambda_{n}(\bx)=\exp\left\{ x^{i}F_{ni}-\psi(\bx) \right\},
\end{eqnarray}
and the square root of $\lambda_{n}$ is the Schmidt coefficient of the wave function
\begin{eqnarray}
\left|\psi\right>=\sum_{n}\sqrt{\lambda_{n}}\left|A;n\right>\otimes\left|\bar{A};n\right>. \label{double}
\end{eqnarray}
The coefficient is normalized as $\left<\psi|\psi\right>=\sum_{n}\lambda_{n}=1$. Usually, the Schmidt coefficient is real, but in this case we generalize this as a complex value. The Fisher metric is given by
\begin{eqnarray}
g_{ij}=\partial_{i}\partial_{j}\psi=\left<F_{i}F_{j}\right>-\left<F_{i}\right>\left<F_{j}\right>, \label{FF}
\end{eqnarray}
where $\left<F_{i}\right>=\sum_{n}\lambda_{n}(t,x)F_{ni}$. In the classical side, we have observed the presence of two copies of a causal cone, and thus it seems better to think that Eq.~(\ref{double}) is represented as
\begin{eqnarray}
\left|\psi\right>=\sum_{n}\sqrt{\lambda_{n}}\left|A;n\right>\otimes\left|\bar{A};n\right>=e^{iHt}\rho^{1/2}\left|I\right>,
\end{eqnarray}
where $\rho$ is the thermal density matrix $\rho=e^{-\beta H}/Z$ for a given Hamiltonian $H$ with inverse temperature $\beta$, $Z={\rm tr}e^{-\beta H}$, and $\left|I\right>$ denotes the identity state $\left|I\right>=\sum_{n}\left|n\right>\otimes\left|\tilde{n}\right>$ where $\left|\tilde{n}\right>=\left|\bar{A};n\right>$ is a basis state in the tilde Hilbert space dual to the original Hilbert space spanned by $\left\{\left|n\right>\right\}$. Therefore, $A$ or $\bar{A}$ can be viewed as a half of the thermofield-double state. The thermofield-double state is an extended notion of spacetime, and this is quite natural generalization of geometric entanglement. 

Note that $F_{n0}$ in Eq.~(\ref{FF}) become necessarily complex if we take the Lorentzian signature. We will discuss this problem later, but we will confirm that there is essentially no problem. We denote $F_{n0}={\rm Re}F_{n0}+2i\phi_{n}$, and the state $\left|\psi\right>$ is then represented as
\begin{eqnarray}
\left|\psi\right>=\sum_{n}e^{i\phi_{n}t}e^{-\frac{1}{2}\beta E_{n}-\frac{1}{2}\ln Z(t,x)}\left|n\right>\otimes\left|\tilde{n}\right>,
\end{eqnarray}
where we can take $H\left|n\right>=\phi_{n}\left|n\right>$ according to the general representation theory $\left|I\right>=\sum_{n}\left|n\right>\otimes\left|\tilde{n}\right>=\sum_{m}\left|m\right>\otimes\left|\tilde{m}\right>$ in Ref.~\cite{Suzuki} and $E_{n}$ is defind by ${\rm Re}F_{n0}$ and $F_{n1}$. We would like to find a relationship between $\beta$ and canonical parameters $\bx=(t,x)$.

\subsection{Mapping from canonical parameters to physical parameters}

Now we are considering a case that $t$ and $x$ are canonical parameters. Thus, this means
\begin{eqnarray}
\lambda_{n}(t,x)=e^{F_{n0}t+F_{n1}x-\psi(t,x)},
\end{eqnarray}
and
\begin{eqnarray}
\left|\psi\right>=\sum_{n}e^{\frac{1}{2}F_{n0}t+\frac{1}{2}F_{n1}x-\frac{1}{2}\psi(t,x)}\left|n\right>\otimes\left|\tilde{n}\right>. \label{TFD2}
\end{eqnarray}
Here the entanglement spectrum is defined by
\begin{eqnarray}
\gamma_{n}(t,x)=-\ln\lambda_{n}(t,x)=\psi(t,x)-F_{n0}t-F_{n1}x.
\end{eqnarray}
Since we have $\left<\partial_{i}\bg\right>=0$ for $i=0,1$, we find
\begin{eqnarray}
\left<\partial_{0}\bg\right> &=& -t-\left<F_{0}\right>=0, \\
\left<\partial_{1}\bg\right> &=& x-\left<F_{1}\right>=0.
\end{eqnarray}
and then
\begin{eqnarray}
\left<F_{0}\right>=-t \; , \; \left<F_{1}\right>=x. \label{f01}
\end{eqnarray}
The entanglement entropy is evaluated as
\begin{eqnarray}
S(t,x)&=&-\sum_{n}\lambda_{n}(t,x)\ln\lambda_{n}(t,x) \nonumber \\
&=&\sum_{n}\lambda_{n}(t,x)\left\{\psi(t,x)-F_{n0}t-F_{n1}x\right\} \nonumber \\
&=& \psi(t,x)-\left<F_{0}\right>t-\left<F_{1}\right>x. \label{entropy}
\end{eqnarray}
Substituting Eq.~(\ref{f01}) into Eq.~(\ref{entropy}) actually leads to Eq.~(\ref{minus}). The imaginary term in $F_{n0}$ does not bring any trouble. Combining Eq.~(\ref{TFD2}) with Eq.~(\ref{entropy}), it is easy to find
\begin{eqnarray}
\left|\psi\right> = \sum_{n}e^{ \frac{1}{2}\left(\delta F_{n0}t+\delta F_{n1}x\right) - \frac{1}{2}S(t,x)}\left|n\right>\otimes\left|\tilde{n}\right>.
\end{eqnarray}
where $\delta F_{n0}$ and $\delta F_{n1}$ are functions of $(t,x)$ since we subtract the $\bx$-dependent averages $\left<F_{0}\right>$ and $\left<F_{1}\right>$ from $F_{n0}$ and $F_{n1}$, respectively. We notice $Z=e^{S}$.

It is important to remember the asymptotic form of $\sqrt{\lambda_{n}}$ for large $n$ in the theory of finite-entanglement scaling~\cite{Lefevre,Pollmann}
\begin{eqnarray}
\sqrt{\lambda_{n}}\sim\exp\left\{-\frac{1}{2}\frac{1}{S}(\ln n)^{2}-\frac{1}{2}S\right\}. \label{CFT}
\end{eqnarray}
This is approximately proved for geometric partition of a CFT in terms of the matrix product state, but it is easy to think that this method can be extended to a case with two copies of the CFT. This consideration suggests
\begin{eqnarray}
S\propto\beta^{-1}=T, \label{ST}
\end{eqnarray}
and we propose
\begin{eqnarray}
\frac{1}{S}E_{n} = -{\rm Re}\left(\delta F_{n0}t + \delta F_{n1}x\right). \label{check}
\end{eqnarray}
The proportionality between $S$ and $T$ is consistent with low-$T$ expansion of the field-theoretical result $S=(c/3)\ln\left((\beta/\pi a)\sinh(\pi x/\beta)\right)$~\cite{Korepin,Calabrese1,Calabrese2}, the finite-$T$ MERA network~\cite{Matsueda3}, and the information-geometrical result~\cite{Matsueda4}. As we have already mentioned, the entropy vanishes at the boundary of the causal surface. This holographic mean is that two Hilbert spaces splits completely at $T=0$ and their quantum correlation vanishes.

To confirm Eq.~(\ref{ST}), we transform $\delta F_{n0}t + \delta F_{n1}x$ into the following form
\begin{eqnarray}
\delta F_{n0}t + \delta F_{n1}x
&=& \left(F_{n0}-\left<F_{0}\right>\right)t+\left(F_{n1}-\left<F_{1}\right>\right)x \nonumber \\
&=& \left(\sum_{m}\lambda_{m}\right)(F_{n0}t+F_{n1}x)-\sum_{m}\lambda_{m}(F_{m0}t+F_{m1}x) \nonumber \\
&=& \sum_{m}e^{\delta F_{m0}t+\delta F_{m1}x-S(t,x)}\left\{\left(F_{n0}-F_{m0}\right)t+\left(F_{n1}-F_{m1}\right)x\right\} \nonumber \\
&=& \frac{e^{-2}}{e^{S(t,x)-1}}\sum_{m}\frac{\left(F_{n0}-F_{m0}\right)t+\left(F_{n1}-F_{m1}\right)x}{e^{-\delta F_{m0}t-\delta F_{m1}x-1}} \nonumber \\
&\simeq& -\frac{e^{-2}}{S(t,x)}\sum_{m}\frac{\left(F_{n0}-F_{m0}\right)t+\left(F_{n1}-F_{m1}\right)x}{\delta F_{m0}t+\delta F_{m1}x},
\end{eqnarray}
and $n$-dependent factor seems to have only weak dependence on the canonical parameters $\bx$. At least for the mean-field approximation $F_{ni}\sim\left<F_{i}\right>$, the $\bx$-dependence on the $n$-dependent factor vanishes. Therefore we can identify it with $E_{n}$, and $S$ is proportional to temperature.

We should emphasize the coordinate transformation from the original model parameters $(t,\beta)$ in the quantum side to canonical parameters $\bx=(t,x)$. In the present case, the temperature in the quantum side is finally converted into the spacetime coordinates in the classical side. Thus, this holography is not simple addition of a scale parameter to the original spacetime coordinates in the quantum side as usually discussed in a context of AdS/CFT. This unique feature has already been discussed in my previous papers~\cite{Matsueda1,Matsueda2}. The feature suggests that the holographic study by the Hessian geometry has much wider functionality than the ordinary holographic theories.

\section{Quantized fields conjugate to canonical parameters}

\subsection{Quantum Fisher metric}

In the above approach, there was an unclear point about complex $\lambda_{n}$ values. We refer to this problem. For this purpose, it is convenient to quantize the classical Fisher metric~\cite{Helstrom,Fujiwara,Petz}. To introduce a quantum Fisher metric, we define the symmetric logarithmic derivative (SLD) Fisher metric as
\begin{eqnarray}
G_{ij}=\frac{1}{2}{\rm tr}\left(\varrho\left\{ \mathcal{L}_{i},\mathcal{L}_{j} \right\}\right), \label{QF}
\end{eqnarray}
where $\varrho$ is the density operator, $\left\{ \mathcal{L}_{i},\mathcal{L}_{j} \right\}=\mathcal{L}_{i}\mathcal{L}_{j}+\mathcal{L}_{j}\mathcal{L}_{i}$, and the SLD operator $\mathcal{L}_{i}$ is defined by a solution of the following linear equation
\begin{eqnarray}
\partial_{i}\varrho=\frac{1}{2}\left\{ \varrho,\mathcal{L}_{i} \right\}=\frac{1}{2}M_{i}\varrho, \label{Liouvillian}
\end{eqnarray}
where $M_{i}=\left\{\ast,\mathcal{L}_{i}\right\}$ is a Liouvillian operator. We easily see that if all the operators commute with each other, Eq.~(\ref{QF}) is equivalent to the classical definition of the Fisher metric. In the commutative case, we find $\partial_{i}(\ln\varrho)=\mathcal{L}_{i}$ and $G_{ij}={\rm tr}\left\{\varrho\left(\partial_{i}\ln\varrho\right)\left(\partial_{j}\ln\varrho\right)\right\}$. We still keep the exponential form of $\varrho$ but $F_{ni}$ are replaced to operators $\mathcal{F}_{i}$ as follows
\begin{eqnarray}
\varrho = \exp\left\{x^{j}\mathcal{F}_{j}-\psi(\bx)\right\}.
\end{eqnarray}
We consider a situation in which $\mathcal{F}_{1}$ is proportional to a unit matrix and $\mathcal{F}_{0}$ has a special matrix form to produce the Lorentzian signature. Note that $\mathcal{F}_{0}$ and $\mathcal{F}_{1}$ are then commutative with each other $\left[\mathcal{F}_{0},\mathcal{F}_{1}\right]=0$.

Let us suppose a pure-state condition $\varrho^{2}=\varrho$. This condition is applicable to the present case $\varrho=\left|\psi\right>\!\left<\psi\right|$ in which we use the thermal vacuum state $\left|\psi\right>$ and its time evolution in TFD. By using the condition as well as comparison with Eq.~(\ref{Liouvillian}), we find
\begin{eqnarray}
\partial_{i}\varrho = \varrho(\partial_{i}\varrho)+(\partial_{i}\varrho)\varrho = \left\{ \varrho , \partial_{i}\varrho \right\}, \label{pure}
\end{eqnarray}
and
\begin{eqnarray}
\mathcal{L}_{i}=2\partial_{i}\varrho.
\end{eqnarray}
The quantum Fisher metric is then represented as
\begin{eqnarray}
G_{ij}=2{\rm tr}\left(\varrho\left\{\partial_{i}\varrho,\partial_{j}\varrho\right\}\right). \label{Gij2}
\end{eqnarray}
If we substitute $\varrho=\left|\psi\right>\!\left<\psi\right|$ into Eq.~(\ref{Gij2}), we obtain the fidelity metric. If we have a direct representation of the wave function, the fidelity metric is also applicable. By taking into account of ${\rm tr}\varrho=1$ and ${\rm tr}\left(\partial_{i}\varrho\right)={\rm tr}\left(\partial_{i}\partial_{j}\varrho\right)=0$, it is straightforward to derive $\partial_{i}\psi={\rm tr}(\varrho\mathcal{F}_{i})$ and
\begin{eqnarray}
\frac{1}{4}G_{ij}={\rm tr}\left(\varrho\mathcal{F}_{i}\mathcal{F}_{j}\right)-{\rm tr}\left(\varrho\mathcal{F}_{i}\right){\rm tr}\left(\varrho\mathcal{F}_{j}\right)=\partial_{i}\partial_{j}\psi=g_{ij}.
\end{eqnarray}
We would like to construct a theory in which we can naturally define $-G_{00}=G_{11}$ and $G_{01}=G_{10}=0$ by chosing $\mathcal{F}_{i}$ appropriately.

\subsection{Time evolution and Lorentzian signature}

For our purpose, it is quite nice to introduce an anti-Hermitian matrix. In this case, pure-imaginary eigenvalues appear, but their complex conjugates also appear. Then, the trace of the density matrix becomes real. Thus, there is no trouble, even if the complex probability appears. Let us consider a simple example with two energy levels given by
\begin{eqnarray}
\mathcal{F}_{0}=\left(\begin{array}{cc}0&-1\\ 1&0\end{array}\right)=-i\sigma_{2} \; , \; \mathcal{F}_{1}=\left(\begin{array}{cc}1&0\\ 0&1\end{array}\right). \label{f0f1}
\end{eqnarray} The eigenvalues of $\mathcal{F}_{0}$ are $\pm i$, and we have ${\rm tr}\left(e^{t\mathcal{F}_{0}}\right)=e^{it}+e^{-it}=2\cos t$. Thus, the sum becomes real. We can simply denote $\mathcal{F}_{0}=-i\mathcal{H}$ with a Hermitian matrix $\mathcal{H}$, and regard the complex eigenvalues as emergence of time evolution by the Hamiltonian $\mathcal{H}$ (time and temperature are mixing with each other when we go to the classical side). Therefore, to describe a previously discussed situation more precisely, the system is in a finite temperature at the initial time, and this evolves in time. It is easy to derive
\begin{eqnarray}
{\rm tr}\left(\varrho \mathcal{F}_{0}\mathcal{F}_{0}\right)-{\rm tr}\left(\varrho \mathcal{F}_{0}\right){\rm tr}\left(\varrho \mathcal{F}_{0}\right)&=&-1 , \\
{\rm tr}\left(\varrho \mathcal{F}_{0}\mathcal{F}_{1}\right)-{\rm tr}\left(\varrho \mathcal{F}_{0}\right){\rm tr}\left(\varrho \mathcal{F}_{1}\right)&=&0 , \\
{\rm tr}\left(\varrho \mathcal{F}_{1}\mathcal{F}_{1}\right)-{\rm tr}\left(\varrho \mathcal{F}_{1}\right){\rm tr}\left(\varrho \mathcal{F}_{1}\right)&=&1 .
\end{eqnarray}
Therefore, Eq.~(\ref{f0f1}) is a good example to understand how the Lorentzian signature appears and how the anti-Hermitian part is related to the time evolution of the thermofield-double state.

\section{Summary}

We have examined the holographic correspondence between causal cones in flat Minkowski spacetime and thermofield double state in the quantum side. We have applied the law of entanglement thermodynamics to derive the entanglement entropy, and have found that the entropy is maximal and is proportional to the temperature in the quantum side. This result is consistent with previous works. At the same time, the positivity of the entanglement entropy is related to the presence of the past and future causal cones in the classical side. We conclude that the Hessian geometrical approach is quite powerful to understand the holographic  connection between quantum field theory and classical spacetime physics. In the present theory based on the Hessian geometry, the temperature in the quantum side was finally mapped onto spacetime coordinates in the classical side. Thus, this holography is more radical transformation than usual AdS/CFT-type correspondence where we just add the radial axis to the spacetime coordinates in the quantum side to construct the classical spacetime. As we have mentioned in my previous two papers~\cite{Matsueda1,Matsueda2}, it is really important to find the mapping of coordinates between both sides.

\acknowledgments
This work was supported by JSPS KAKENHI Grant Number 15K05222.

\end{document}